\documentclass[10pt,conference]{IEEEtran}

\usepackage[a-1b]{pdfx}

\usepackage{algorithm2e}
\usepackage{booktabs}
\usepackage{enumitem}
\usepackage{epigraph}
\usepackage{graphicx}
\usepackage[utf8]{inputenc}
\usepackage[frozencache=true]{minted}
\usepackage{subcaption}
\usepackage{tabularx}
\usepackage[para,online,flushleft]{threeparttable}
\usepackage{url}
\usepackage{xspace}
\usepackage{zi4}

\newcommand\insight[1]{
    \vspace{0.25cm}
	\noindent 
	\fcolorbox{gray!20}{gray!20}{
		\parbox{0.92\columnwidth}
		{#1}
		\hspace*{0.5ex}
	}
}

\makeatletter
\newcommand*{\rom}[1]{\expandafter\@slowromancap\romannumeral #1@}
\makeatother

\def\Tool{V2}
\def\Search{DriftSearch}
\def\feedback{feedback-directed search}
\def\Feedback{Feedback-directed search}
\def\FEEDBACK{Feedback-Directed Search}
\def\mobi{\texttt{moby.py}}

\usepackage{amsfonts}
\usepackage{amsmath}
\usepackage{tikz}

\usetikzlibrary{calc}

\newenvironment{tightcenter}
    {\parskip=0pt\par\nopagebreak\centering}
    {\par\noindent\ignorespacesafterend}

\newlength{\RoundedBoxWidth}
\newsavebox{\GrayRoundedBox}
\newenvironment{GrayBox}[1]{
    \setlength{\RoundedBoxWidth}{\linewidth-4.5ex}
    \def\boxheading{#1}
    \begin{lrbox}{\GrayRoundedBox}
        \begin{minipage}{\RoundedBoxWidth}
}{
        \end{minipage}
    \end{lrbox}
    \begin{tightcenter}
        \begin{tikzpicture}
            \node(Text)[draw=black!20,fill=white,rounded corners,inner sep=2ex,text width=\RoundedBoxWidth] {\usebox{\GrayRoundedBox}};
            \coordinate(x) at (current bounding box.north west);
            \node [draw=white,rectangle,inner sep=3pt,anchor=north west,fill=white] at ($(x)+(10.5pt,.75em)$) {\boxheading};
        \end{tikzpicture}
    \end{tightcenter}
    \vspace{0pt}
    \ignorespacesafterend
}

\newenvironment{problem}[1]{
    \noindent\ignorespaces
    \FrameSep=6pt
    \parindent=0pt
    \vspace*{-.5em}
    \begin{GrayBox}{\textsc{#1}}
        \newcommand\Prob{Problem:}
        \newcommand\Input{Input:}
        
        \newcommand\Objectives{Objectives:}
        \begin{tabular*}{\columnwidth}{@{\hspace{.5em}} >{\itshape} p{1.4cm} p{0.85\columnwidth} @{}}
}{
        \end{tabular*}
    \end{GrayBox}
    \vspace*{-.5em}
    \ignorespacesafterend
}

\setminted{style=borland,xleftmargin=15px,linenos,firstnumber=1,fontsize=\small,autogobble,breaklines}

\makeatletter
\renewcommand{\@endalgocfline}{\relax}
\makeatother
\SetKwProg{Procedure}{Procedure}{}{}
\SetKwRepeat{DoWhile}{do}{while}

\title{V2: Fast Detection of Configuration Drift in Python}

\author{
    \IEEEauthorblockN{Eric Horton, Chris Parnin}
    \IEEEauthorblockA{
        North Carolina State University\\
        Raleigh, NC, USA\\
        \{ewhorton, cjparnin\}@ncsu.edu
    }
}

\begin{document}

    \maketitle

    \begin{abstract}

        Code snippets are prevalent, but are hard to reuse because they often lack an accompanying environment configuration. Most are not actively maintained, allowing for drift between the most recent possible configuration and the code snippet as the snippet becomes out-of-date over time. Recent work has identified the problem of validating and detecting out-of-date code snippets as the most important consideration for code reuse. However, determining if a snippet is correct, but simply out-of-date, is a non-trivial task. In the best case, breaking changes are well documented, allowing developers to manually determine when a code snippet contains an out-of-date API usage. In the worst case, determining if and when a breaking change was made requires an exhaustive search through previous dependency versions.

        We present \Tool{}, a strategy for determining if a code snippet is out-of-date by detecting discrete instances of configuration drift, where the snippet uses an API which has since undergone a breaking change. Each instance of configuration drift is classified by a failure encountered during validation and a configuration patch, consisting of dependency version changes, which fixes the underlying fault. \Tool{} uses \feedback{} to explore the possible configuration space for a code snippet, reducing the number of potential environment configurations that need to be validated. When run on a corpus of public Python snippets from prior research, \Tool{} identifies 248 instances of configuration drift.

    \end{abstract}



    \section{Introduction}{

        Code snippets are commonly used by developers to provide API documentation~\cite{Parnin:blogs:2013} and act as examples for learning and reuse~\cite{Wang:2015, Horton:Gistable:2018, Rule:2018}. Stack Overflow, a question and answer site for developers, has over 1M questions relating to Python, a popular and fast growing programming language~\cite{Wang:2015}. On GitHub's gist system, developers have shared over 300K public Python code snippets~\cite{Horton:Gistable:2018}. Both feature prominently in search results for API documentation~\cite{Treude:2018}, and a study by Yang et al. found over 4M code blocks from Stack Overflow snippets that had been reused in public GitHub projects~\cite{Yang:2017}. Recently, code snippets in Jupyter notebooks have become a standard for sharing and replicating scientific work and more~\cite{Rule:2018}.

        Unfortunately, many code snippets require a non-trivial environment configuration in order to execute successfully~\cite{Yang:2016, Sulir:2016}, and are not accompanied by sufficient information for developers to easily recreate that configuration~\cite{Horton:Gistable:2018, Horton:2019:DockerizeMe, Pimentel:2019:Notebooks}. This leads to the problem of configuration drift, where a code snippet goes out-of-date because the APIs that it depends on experience breaking changes over time. Developers struggle to determine if a code snippet has experienced configuration drift, or is simply incorrect~\cite{Horton:Gistable:2018}. In some instances, a breaking change made to an API may be highlighted in release documents made available to developers. In the worst case, determining if and when a breaking change was made requires an exhaustive search through previous dependency versions.

        The ability to validate and detect out-of-date code snippets is the most important consideration for quality of code reuse according to a recent survey of 183 software developers~\cite{Wu:2018:SOReUse}. Pimental et al. found that only 24\% of Jupyter notebooks could be executed, and only 4\% had reproducible results~\cite{Pimentel:2019:Notebooks}. They note that reproducibility suffers because the notebook format does not encode dependencies or dependency versions. For example, the recent release of Tensorflow version 2.0 introduced many breaking changes. As a result, many Jupyter notebooks are not runnable with the latest version of the framework. This proves detrimental, as developers have remarked that documentation for the Tensorflow platform is largely based on examples.\footnote{https://news.ycombinator.com/item?id=18445225} Overall, there is an unmet need for checking if code examples are up-to-date and runnable.

        This work presents \Tool{}, available at {\small \textbf{\url{https://github.com/v2-project/v2}}}, a tool that determines if a code snippet is out-of-date by detecting configuration drift. \Tool{} is based on the observation that an instance of configuration drift often manifests as a crash during execution. It can therefore be categorized by the illuminating failure (crash) paired with a configuration patch sufficient to enable execution. Patches consist of changes to dependency versions, and act both as a certificate of correctness for the snippet and enable execution with an older configuration if desired.

        Unlike other work in configuration repair~\cite{Macho:2018:BuildMedic, Hassan:HireBuild}, \Tool{} automatically infers up-to-date candidate environment configurations from a code snippet without the need of developer input or pre-existing build scripts. If the code snippet experiences a crash when executed in its candidate environments, \Tool{} searches for configuration drift by employing \textit{\feedback{}}, a search strategy that incorporates feedback from code snippet execution and knowledge about prior API breakage events to prune and prioritize configuration patches from the space of all possible environment configurations.

        We show that \Tool{} is able to discover configuration drift in open source Python code snippets from the Gistable dataset~\cite{Horton:Gistable:2018} and from a wide array of Jupyter notebooks~\cite{Rule:2018}. For both datasets, we show that incorporating execution feedback and knowledge of previous API breakages enables \Tool{} to more quickly decide when configuration drift is present in a snippet.

        \begin{figure*}

    \centering

    \begin{subfigure}[t]{0.45\textwidth}
        \centering
        \begin{minted}{Python}
            import math
            from functools import wraps
            
            from theano import tensor as T
            from theano.sandbox.rng_mrg import MRG_RandomStreams as RandomStreams
            
            from lasagne import init
            from lasagne.random import get_rng
            
            ...
        \end{minted}
        \caption{https://gist.github.com/a003ace716c278ab87669f2fbd37727b}
        \label{fig:motivation:snippet}
    \end{subfigure}
    \hspace{1cm}
    \begin{minipage}[t]{0.45\textwidth}
        \begin{subfigure}[t]{\textwidth}
            \centering
            \begin{minted}{Dockerfile}

                FROM python:3.7
                ADD bayes.py /bayes.py
                RUN ["pip", "install", "Theano==1.0.4"]
                RUN ["pip", "install", "Lasagne=0.1"]
                CMD python /bayes.py

            \end{minted}
            \caption{Dockerfile}
            \label{fig:motivation:dockerfile}
        \end{subfigure}

        \vspace{0.75cm}

        \begin{subfigure}[t]{\textwidth}
            \centering
            \begin{minted}{Text}

               DecrementSemverMajor(Theano==1.0.4) -> 0.9.0
               DecrementSemverMinor(Theano==0.9.0) -> 0.8.2

            \end{minted}
            \caption{Mutations}
            \label{fig:motivation:mutations}
        \end{subfigure}        
    \end{minipage}

    \caption{
        (a) Import statements extracted from a GitHub gist using the Lasagne and Theano APIs.
        (b) A base environment configuration, using the Docker container system Dockerfile format, containing the gist and the latest versions of Theano and Lasagne. Due to configuration drift, the gist does not execute successfully using the configured versions.
        (c) Environment mutations made while searching for the correct version of Theano.
    }
    \label{fig:dashtable}

\end{figure*}

    }

    \section{Motivation}{

        Consider the code snippet fragment presented in Figure~\ref{fig:motivation:snippet}. The fragment contains the import statements for \texttt{Lasagne} and \texttt{Theano} from a public Python gist found on GitHub. Both libraries can be found on and installed from the Python Package Index (PyPI). However, after installing version \texttt{0.1} and \texttt{1.0.4} (the latest versions), respectively, executing the code fragment with Python 3 results in the following exception:

        \begin{minted}[breakafter=/]{Text}

            ...
            File "/usr/local/lib/python3.7/site-packages/lasagne/layers/pool.py", line 6, in <module>
              from theano.tensor.signal import downsample
            ImportError: cannot import name 'downsample' from 'theano.tensor.signal' (/usr/local/lib/python3.7/site-packages/theano/tensor/signal/__init__.py)

        \end{minted}

        The exception indicates Lasagne has attempted to import \texttt{downsample} from Theano and failed because the current version of Theano does not provide it in the expected location. The Lasagne installation documentation\footnote{https://lasagne.readthedocs.io/en/latest/user/installation.html} indicates that the library is tightly coupled with Theano, and that a very recent version of Theano is often required to run correctly:

        \begin{quote}
            Lasagne has a couple of prerequisites that need to be installed first, but it is not very picky about versions. The single exception is Theano: Due to its tight coupling to Theano, you will have to install a recent version of Theano (usually more recent than the latest official release!) fitting the version of Lasagne you choose to install.
        \end{quote}

        However, the current version of Lasagne was released before the current version of Theano. While the documentation indicates that the correct configuration requires installing a pre-release version of Theano, it may actually be the case that Lasagne has drifted and is now out-of-date. The Theano versioning scheme uses semantic versioning\footnote{https://semver.org/} (semver), which tells us the current version of Theano is a major version with some patches. We start searching for the correct version by going back to the latest release of the previous major version, \texttt{Theano==0.9.0}. When that fails, we go back an additional minor version, \texttt{Theano==0.8.2}. Using this version allows the gist to be executed successfully, although with a warning that the ``downsample module has been moved to the theano.tensor.signal.pool module,'' confirming that execution originally failed because the code had fallen out-of-date.

    }

    \section{\Tool{}}\label{sec:v2}{

        \Tool{}, \emph{``Version 2''}, is an extension of DockerizeMe~\cite{Horton:2019:DockerizeMe} that adds support for validating multiple environments and reasoning over dependency versions. \Tool{} determines if a code snippet or its dependencies is out-of-date by discovering instances of configuration drift. Each instance of configuration drift is identified by two things: 1) a runtime failure and 2) a configuration patch consisting of version changes.

        To detect configuration drift, \Tool{} first generates candidate environments that are potentially correct. That is, the environment is configured with the dependencies that \Tool{} infers as being potentially necessary for executing a snippet correctly. It then executes the snippet in this inferred environment, recording any execution failures as they are encountered. For each new execution failure, \Tool{} applies mutations to installed dependency versions to generate a new candidate configuration until the fault is either resolved, or no new candidate configurations can be made. \Tool{} uses information about snippet execution and prior API breakage events to prune and prioritize candidate configurations. We now detail each stage of the repair algorithm in detail.


        \subsection{Candidate Environment Generation}\label{sec:v2:candidate-generation}{

            In order to find instances where configuration drift has caused a failure, \Tool{} must first have one or more candidate environment configurations for a code snippet. These should contain the set of required dependencies, although they may not have the correct versions. \Tool{} uses the Docker container system to specify configurations for isolated environments. The use of Docker guarantees a clean and consistent base environment for \Tool{} to work in, and allows each candidate environment to be based off of a standard image.

            Although the scheduled end-of-life for Python 2 is 2020~\cite{PEP:373}, many older scripts will not run in Python 3, as they use syntax that was removed from the language. Candidate generation must first determine which language runtime to use: either Python 2 or Python 3. The correct language runtime is detected by attempting to parse the code snippet using Python's built in AST module for each language runtime (2 or 3). If the code snippet parses in both language runtimes, then two potential candidate environments are generated.

            \Tool{} then uses the dependency resolution algorithm and knowledge base provided by DockerizeMe~\cite{Horton:2019:DockerizeMe} to populate each candidate environment with an initial set of dependencies in the following manner. \Tool{} extracts fully qualified names of resources appearing in import nodes of the AST. Imported resources not in the Python standard library are mapped to a set of packages from the Python Package Index (PyPI) that potentially provide them, although the set may be an overestimate. These packages are considered the direct dependencies of the code snippet. \Tool{} also resolves all transitive dependencies, i.e., packages depended on by a direct dependency or by another transitive dependency. Each dependency is initially pinned at the latest version available on PyPI, because DockerizeMe does not consider versions, and a final installation order is generated such that, for each dependency, it is installed only after all packages that it depends on.

            A \( (runtime, dependencies) \) tuple defines a candidate environment specification containing the correct language runtime level and dependencies listed in a valid installation order.

        }

        \subsection{Environment Validation}\label{sec:v2:validation}{

            The \Tool{} validation phase accepts as input a single code snippet and a candidate environment specification in the \( (runtime, dependencies) \) format. It returns a status code indicating the result of executing the code snippet in the candidate environment, plus any metadata about the execution or encountered failures. Validation consists of two phases: 1) environment configuration and 2) snippet execution.

            During the configuration step, \Tool{} creates a new execution environment using a Docker container based on the language runtime. This guarantees a consistent starting environment. It then configures the execution environment according to the candidate environment specification by installing each dependency in order. Installation failures are recorded as part of the validation metadata, but do not halt validation. If a dependency which is not required fails to install, it will not affect the result of execution. Conversely, if a required dependency fails to install, it will produce an execution failure.

            Finally, the snippet is executed in the configured environment. If the snippet runs to completion without experiencing a failure, the execution result is coded as \texttt{Success}. If an exception is encountered during execution, the snippet is considered to have failed validation and the execution result is coded as \texttt{Exception}. In this case, the exception name, message, and stack trace are provided in the validation metadata.

            In certain cases, a code snippet will run indefinitely instead of exiting. For example, a snippet which waits for user input will block forever. If a snippet has not exited after a time limit, execution is stopped and the validation is coded as \texttt{Timeout}. A timeout is considered neither a success nor a failure, as it is generally undecidable if the snippet would have eventually exited. Regular Python snippets are run with a time limit of one minute. Because they generally involve computation and require a longer runtime, Jupyter notebooks are run with a base time limit of two minutes, plus an additional minute for each cell in the notebook.

        }

        \subsection{Environment Mutation}\label{sec:v2:mutation}{

            Configuration drift is classified by a validation failure and a patch. Each patch consists of a sequence of configuration changes (mutations) to the environment configuration which are sufficient to resolve the fault. \Tool{} supports two classes of mutation operators for dependency versions. The first class is based on the semantic versioning scheme, and the second class is based on pre-existing knowledge about version changes.

            \subsubsection{Semantic Versioning}\label{sec:v2:mutation:semver}{

                Semantic versioning (semver) is a versioning scheme which defines major, minor, patch, and prerelease changes to a package. Breaking API changes must be accompanied by a major version change, while backwards compatible additions are accompanied by a minor version change. However, before version \texttt{1.0.0}, semver defines that an API should not be considered stable, and that anything may change at any time. Many Python dependencies follow a versioning scheme that is, or can be interpreted as, semver.\footnote{https://www.python.org/dev/peps/pep-0440/}

                \Tool{} defines two mutation operators for semantic versions: 1) \texttt{DecrementSemverMajor} and 2) \texttt{DecrementSemverMinor}, which mutate the major and minor version level of a dependency in an environment configuration as seen in Figure~\ref{fig:motivation:mutations}. Both mutation operators will decrement the specified version level by one if possible, but will choose the latest release at that version level. For example, if a package has versions \texttt{2.1.0}, \texttt{1.2.0}, and \texttt{1.1.0}, then \texttt{DecrementSemverMajor(2.1.0) -> 1.2.0}. \texttt{DecrementSemverMinor} only operates within a major version, it will not roll back to a previous minor version if doing so also requires decrementing the major version.

            }

            \subsubsection{Upgrade Matrix}{

                \begin{figure*}
    \centering
    \begin{subfigure}[t]{0.45\textwidth}
        \centering
        \includegraphics[width=\linewidth]{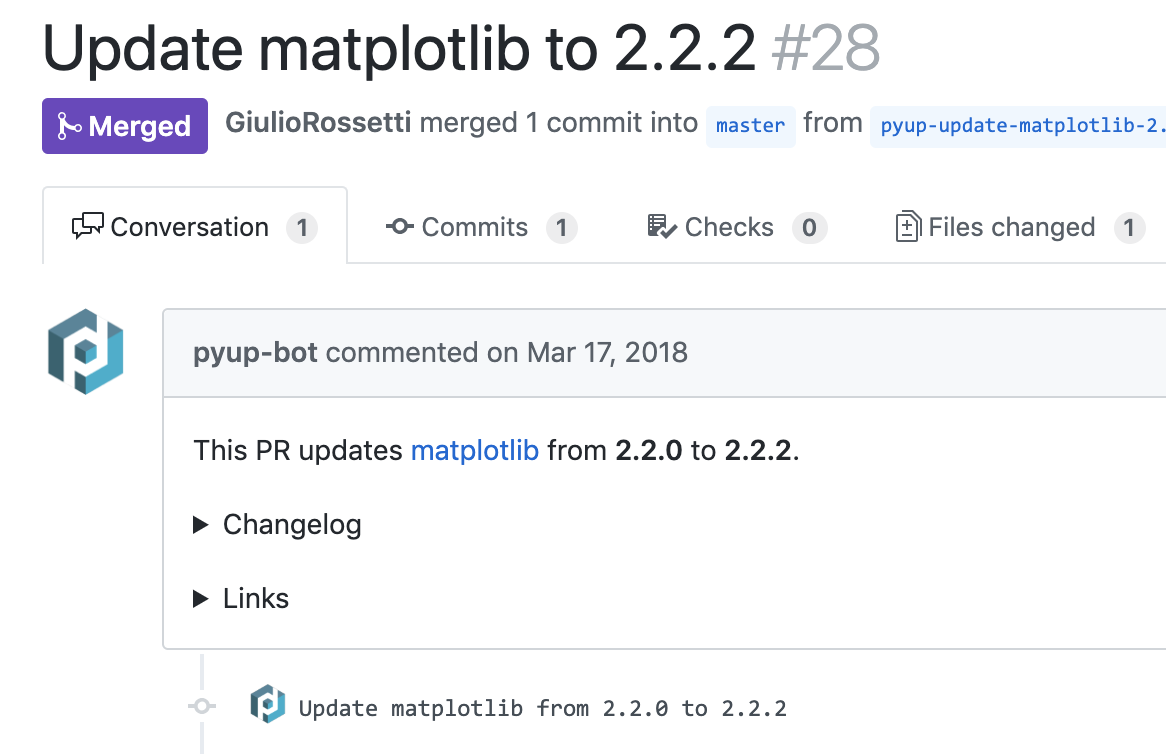}
        \caption{Pull request to upgrade matplotlib from 2.2.0 to 2.2.2.}
        \label{fig:upgrade-bot:pull-request}
    \end{subfigure}
    \hspace{1cm}
    \begin{subfigure}[t]{0.45\textwidth}
        \centering
        \includegraphics[width=\linewidth]{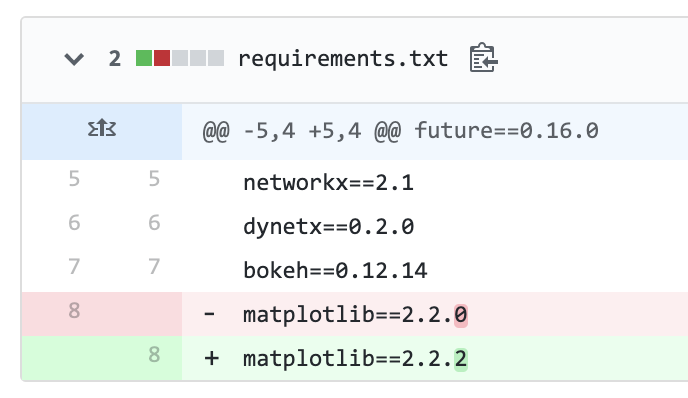}
        \caption{Git diff.}
        \label{fig:upgrade-bot:diff}
    \end{subfigure}
    \caption{An upgrade event initiated by the upgrade bot PyUp.}
    \label{fig:upgrade-bot}
\end{figure*}

                \begin{figure}[!ht]
    \centering
    \includegraphics[width=\linewidth]{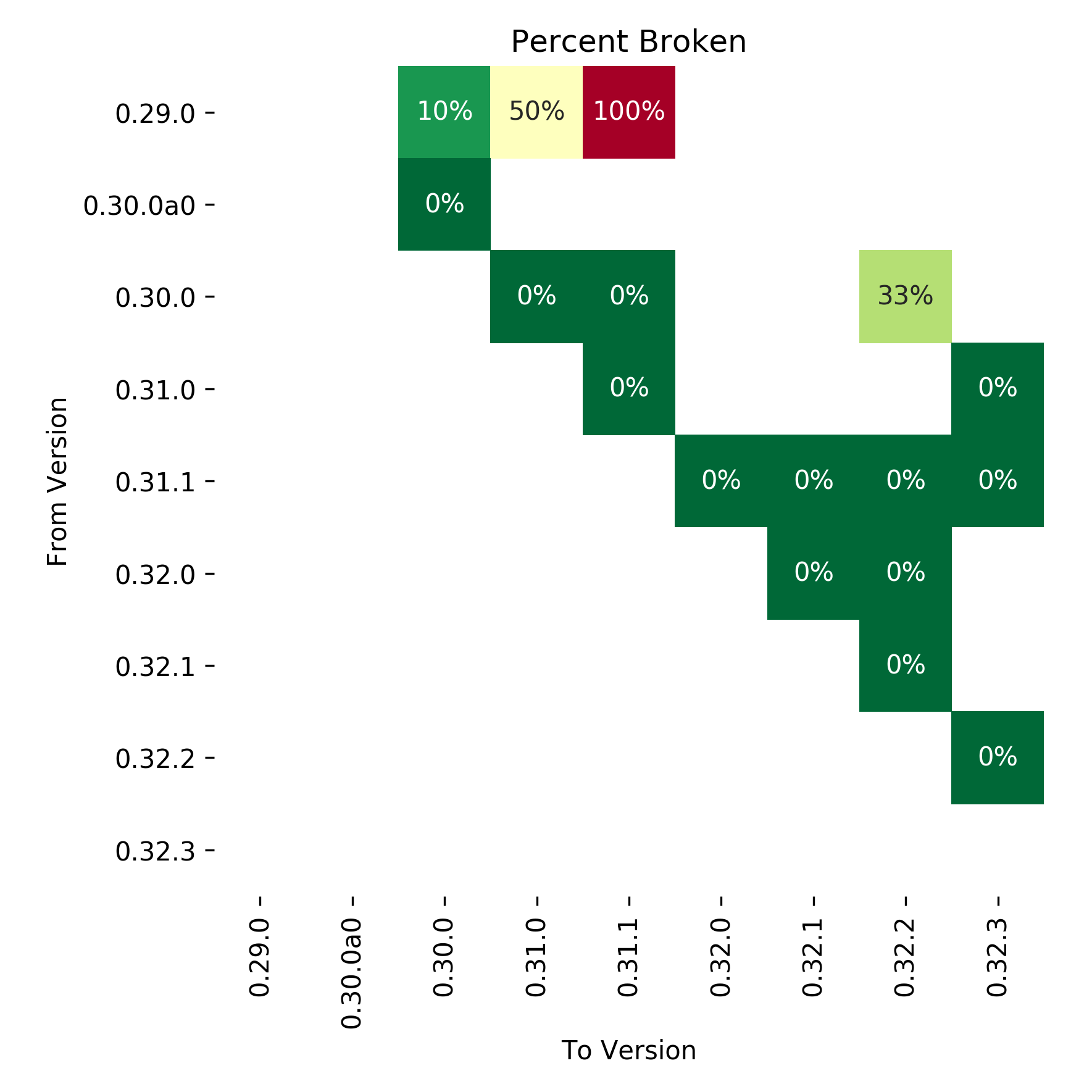}
    \caption{Version upgrade matrix for Wheel.}
    \label{fig:wheel-upgrade-matrix}
\end{figure}

                In the worst case, all versions of a dependency may be enumerated by combining the semver mutation operators described in Section~\ref{sec:v2:mutation:semver}. However, such a brute force approach is generally not feasible due to the large number of potential candidate environments in the version configuration space. A more efficient approach is to prioritize or prune environment configurations based on how likely they are to fix a validation failure.

                To make this estimation, we extrapolate from TravisCI build statuses of Python projects that experience version upgrade events. An upgrade event occurs when a commit is made that upgrades the version of a single project dependency, triggering a TravisCI build for the project. Figure~\ref{fig:upgrade-bot} shows an upgrade event initiated by \texttt{pyup-bot}, an upgrade bot for Python projects.

                All upgrade events were discovered by mining Google BigQuery for GitHub pull requests from January 2014 though January 2019. We limit our search to pull requests which change the version for only a single Python dependency and were submitted by an account with the name \texttt{dependabot[bot]}, \texttt{dependabot}, \texttt{snyk-bot}, or \texttt{pyup-bot}. We then extract the project build statuses for the original and new dependency versions from Travis CI, excluding build histories where one of the statuses was either canceled or errored. In total, this provided 7,104 upgrade events for 193 distinct packages that experienced at least one failure.


                Taken together, the build statuses for a single package form a version upgrade matrix for that package, where each row and column indicates upgrading from one version to another. Each cell of an upgrade matrix contains the percent of builds that were broken by performing the version upgrade, where a breakage is determined by a build status that changes from passing to failing as a result of the upgrade. Figure~\ref{fig:wheel-upgrade-matrix} shows the version upgrade matrix for the Python package Wheel. An incompatibility cluster in the top row indicates that upgrading from version \texttt{0.29.0} becomes more likely to cause a failure as the distance between versions increases.

                Intuitively, we apply the heuristic that upgrades which break a larger number of builds are more likely to indicate a backwards incompatible change. Conversely, given a failed validation, downgrading to a version which experienced a large number of upgrade failures is more likely to fix the fault. When \Tool{} finds a validation failure caused by a dependency with an upgrade matrix, it can leverage information about the incompatible upgrades to test only those versions likely to result in a fix, pruning the rest of the version space.

            }

        }

        \subsection{\FEEDBACK{}}\label{sec:v2:feedback}{

            The problem of finding an instance of configuration drift can be solved by conducting a search through the full configuration space for a code snippet. We model the configuration space as a configuration graph \(G = (V, E)\), where each vertex \(v \in V\) is a potential environment configuration and each directed edge \((u, v, m) \in E\) is a pair of configurations \( (u, v) \) and a mutation \( m \) such that applying mutation \( m \) to configuration \( u \) results in configuration \( v \). We now formally define the problem, \Search{}, of finding patched configurations in \( G \) that fix faults caused by configuration drift. If some configuration fixes all faults, we say that it is a working environment configuration.

            \begin{problem}{\Search{}}
                \Input      & A configuration graph \( G = (V, E) \) and a   \\
                            & starting environment \( v \).                  \\
                \Prob       & Find a working environment \( w \).            \\
                \Objectives & Minimize the distance from \( v \) to \( w \). \\
                            & Minimize the number of vertices explored.      \\
            \end{problem}

            We implement a search strategy, \textit{\feedback{}}, that generates a configuration space from a candidate environment configuration by applying the mutation operators defined in Section~\ref{sec:v2:mutation}. \Feedback{} incorporates information about the executability of a code snippet within candidate environments to intelligently drive exploration of the configuration space, reducing the number of environments in the configuration space explored and allowing \Tool{} to quickly converge to a working configuration.

            \begin{algorithm}
                \Procedure{FeedbackDirectedSearch(environment)}{
                    checkpoint = null\;
                    \DoWhile{not successful(checkpoint)}{
                        \tcp{Validation}
                        validation = validate(environment)\;
                        \If{fixed(checkpoint, validation)}{
                            record\_drift(checkpoint)\;
                            checkpoint = validation\;
                            \If{not fixable(checkpoint)}{
                                \KwRet environment\;
                            }
                        }
                        \;
                        \tcp{Fault Localization}
                        \If{updated(checkpoint)}{
                            dependency = localize(checkpoint)\;
                            \If{dependency is not null}{
                                \If{has\_matrix(dependency)}{
                                    mutator = matrix(dependency)\;
                                }
                                \Else{
                                    mutator = IDDFS(dependency)\;
                                }
                            }
                            \Else{
                                mutator = IDDFS(environment)\;
                            }
                        }
                        \;
                        \tcp{Exploration}
                        environment = next(environment, mutator)\;
                    }
                    \KwRet environment\;
                }
                \vspace{0.25cm}
                \caption{
                    Implementation of \feedback{} for a single candidate environment. Exploration is performed by either using a version upgrade matrix or iterative-deepening depth-first search (IDDFS).
                }
                \label{alg:feedback-directed}
            \end{algorithm}

            \Tool{} begins \feedback{} with a list of candidate environment configurations generated as described in Section~\ref{sec:v2:candidate-generation}. If more than one candidate environment was generated, \feedback{} operates on the environment specifications round-robin, exploring the configuration space for each candidate in tandem. Doing so allows \feedback{} to satisfy the objective of minimizing the distance, or number of mutations, from starting environment \( v \) to working environment \( w \) in the case that one candidate environment is easily fixable, but the other is not. The search algorithm is structured in three distinct phases: validation checkpointing and pruning, fault localization, and exploration. A high-level implementation of \feedback{} for a single environment is outlined in Algorithm~\ref{alg:feedback-directed}. We now highlight each phase in detail.

            \subsubsection{Validation Checkpointing and Pruning}\label{sec:v2:feedback:checkpointing}{

                Given a code snippet \( s \) and candidate environment \( c \), \feedback{} first performs validation of \( s \) in \( c \) using the procedure outlined in Section~\ref{sec:v2:validation}. If the candidate environment produces an execution failure, the validation result is saved as a checkpoint, and \( c \) will be mutated to fix the fault. After each environment mutation, \( s \) is again validated in \( c \).

                Conceptually, at each stage of the search process, a validation checkpoint represents the latest unfixed validation failure. We consider the fault indicated by the validation checkpoint to have been fixed by a sequence of mutations to \( c \) if a newer validation results in an execution which covers more of \( s \). That is, the mutations made to \( c \) allow execution of \( s \) to proceed past the line which previously caused a failure. Whenever such a sequence is discovered by \feedback{}, it, and the validation checkpoint, are recorded as an instance of configuration drift, and the checkpoint is updated to the newer validation. If a validation indicates that a sequence of mutations has not changed the validation result, or that execution exits on or before the line reached by the checkpoint, it is discarded and the search process continues.

                Whenever a new validation checkpoint is discovered, \Tool{} determines if it is potentially fixable via mutations to dependency versions. If it is not potentially fixable, search halts and returns the current environment and instances of configuration drift reported. A validation checkpoint is considered potentially fixable if the execution exception does not indicate a failure due to the local file system and satisfies one of:

                \begin{enumerate}
                    \item It is caused by an installed dependency.
                    \item It is an import error related to an installed dependency.
                    \item It is one of \texttt{TypeError} or \texttt{AttributeError}, which can indicate breaking changes in a public API.
                \end{enumerate}

            }

            \subsubsection{Fault Localization} {

                During search, \Tool{} prunes the local configuration search space by performing fault localization to map validation checkpoint failures to a single dependency. If the execution failure was caused by code not belonging to the code snippet or the Python standard library, \Tool{} extracts the last dependency from the stack trace that was installed as part of the environment configuration. If the exception is an import error related to a single installed dependency, \Tool{} indicates that dependency. In cases where fault localization fails to highlight a single installed dependency related to the failure, \Tool{} continues exploring the local search space.

            }

            \subsubsection{Exploration}{

                Whenever a validation result indicates that a code snippet \( s \) does not execute successfully in a candidate environment \( c \), \feedback{} searches the local configuration space for a fix by applying mutations to \( c \). There are three exploration strategies, based on the success of fault localization and whether \Tool{} can leverage one or more version upgrade matrices.

                If \Tool{} succeeds in localizing a fault to a single installed dependency \( d \), and that dependency has a version upgrade matrix, \feedback{} proceeds by querying all pairs of versions \( (v_{i, 1}, v_{i, 2}) \) such that the version matrix indicates a nonzero percent of builds were broken by upgrading from \( v_{i, 2} \) to \( v_{i, 1} \), and \( v_{i, 1} \) is at most the current version of \( d \). Selection in this manner implicitly disregards the versions which experienced no breaking upgrades. An ordering of versions is generated by sorting all pairs in descending order by \( v_{i, 1} \). Then, for each \( v_{i, 1} \), all \( v_{j, 2} \) not already in the ordering are sorted in descending order by percentage of builds broken and appended to the order. While validation indicates that the checkpoint has not been fixed, \( d \) is mutated to be the next version in the ordering and validated again.

                When \Tool{} is capable of localizing a fault to a single installed dependency \( d \), but that dependency does not have a version upgrade matrix, \feedback{} explores the local configuration space by performing iterative-deepening depth-first search (IDDFS) over the versions of \( d \) using the semver mutation operators from Section~\ref{sec:v2:mutation:semver}. In all cases, the operators are strongly ordered such that all instances of \texttt{DecrementSemverMajor} are applied prior to instances of \texttt{DecrementSemverMinor}. This, coupled with the fact that \texttt{DecrementSemverMinor} does not cross major version boundaries, guarantees that the same configuration is not reached at two different depths. If \Tool{} is unable to localize a fault to a single installed dependency, \feedback{} performs iterative-deepening depth-first search over the versions of all dependencies, making use of upgrade matrices for individual dependencies where applicable. In either case, while performing iterative-deepening depth-first search, candidate environments are only validated when they are generated.

                Because of the size of the search space, tracking all candidate environments on the frontier quickly becomes intractable, limiting search algorithms which require doing so. Depth-first exploration allows \feedback{} to generate and validate new candidate environments with smaller memory requirements. Note that, because mutation operators result in decreasing dependency versions, exploration will never undo mutations made prior to the current checkpoint. Search is halted if no sequence of mutations will lead to an as-of-yet unexplored environment.

            }

        }

    }

    \section{Evaluation}{

        We evaluate the ability of \Tool{} to find instances of configuration drift by analyzing open source Python code snippets.

        \subsection{Datasets}{

            \begin{table}
                \caption{Summary of code snippets from datasets.}
                \label{tbl:dataset-summary}
                \centering
                \begin{tabular}{l|rr}
                    \toprule
                                                                            & \textbf{Gist} & \textbf{Jupyter} \\
                    \midrule
                    Total                                   & 2,096          & 6,529 \\
                    With a Candidate Environment            & 2,096          & 5,423 \\
                    With a Successful Candidate Environment &   119          &   758 \\
                    No Successful Candidate Environment     & 1,977          & 4,665 \\
                    \bottomrule

                \end{tabular}
            \end{table}

            Horton and Parnin previously presented Gistable, a dataset of 10,259 Python code snippets from GitHub's snippet sharing service~\cite{Horton:Gistable:2018}. They identified a subset of approximately 2,891 ``hard'' gists that experienced a failure even after a straightforward approach to inferring an environment configuration. We exclude 795 gists that are known to have over 10 direct dependencies, because the large configuration space is intractable for a baseline. For example, an environment with 10 dependencies, each with 3 versions, would have \( 3^{10} = \text{nearly 60K} \) possible unique configurations. This leaves 2,096 gists that we refer to as the Gist dataset. Prior work was unable to make 1,999 of these gists execute successfully after candidate environment generation using the latest versions of all dependencies~\cite{Horton:2019:DockerizeMe}.

            Second, we collect larger Python code snippets in the form of Python notebooks. Rule et al. previously collected 1.25M open source Jupyter notebooks from GitHub, making the dataset publicly available~\cite{Rule:2018}. As part of their release, they provide a random sample of 6,529 notebooks, many of which are written in Python. We exclude 1,106 notebooks from the sample for which no candidate environment could be generated, leaving 5,423 notebooks we refer to as the Jupyter dataset. There were several main causes for why candidate generation failed, all related to being unable to parse the snippet. Some notebook files were completely empty, or identified as using Jupyter kernels for other languages, such as R, meaning \Tool{} could not parse the notebook source code as Python. Other notebooks contained IPython magics,\footnote{\url{https://ipython.readthedocs.io/en/stable/interactive/magics.html}} which are defined as statements which are syntactically invalid Python, causing difficulty in parsing.


            For simplicity, and to disambiguate between the origin of code snippets, we refer to code snippets from the Gist dataset as ``gists'' and code snippets from the Jupyter dataset as ``notebooks.'' Table~\ref{tbl:dataset-summary} shows a summary of both. Together, these snippets represent a wide range of behaviors and API uses, and their generated candidate environments have over 2K unique installable packages.

        }

        \subsection{Methodology}{

            We use iterative-deepening depth-first search to establish a baseline for identifying configuration drift in each dataset. This baseline enumerates all states in the configuration space and represents a worst-case scenario where it is assumed a repair can be made by modifying a dependency version, but no information is available to guide configuration. For each code snippet, we record whether search finished or timed out. If search finished, we record whether or not it resulted in a successful environment configuration, and the total number of environments validated.

            After running the baseline, we analyze both datasets using \Tool{} with \feedback{}, recording the same metrics as used in the baseline evaluation. In all cases where search finished within the timeout, we also record every instance of configuration drift identified by \Tool{}, along with the number of environments in the configuration space that were validated while identifying the instance, as evaluations are the dominating factor in cost for validation based approaches~\cite{Weimer:2013:Repair}. How far the final configuration is from the starting configuration is also recorded. If search did not find a fully working environment configuration, we record metadata about why the search process terminated, and which portion of the configuration space it was operating in.

            We performed evaluation on a cluster of 8 virtual compute nodes running Ubuntu 16.04.5 LTS. Each compute node was configured with 8 CPU cores and 12 GB memory. Evaluations were run using the HashiCorp Nomad job scheduler, with at most 6 jobs running on a single compute node at one time. All evaluations were run with a timeout of 1 hour, after which they were given the chance to gracefully exit and record results. Evaluations which lasted longer than 2 hours were terminated. This could happen for snippets where validating the last environment took longer than an hour, as we allowed each validation to fully complete before timing out.

        }

    }

    \section{Results}{

        We evaluate \Tool{} with three metrics related to its ability to detect configuration drift: improvement over the baseline, discovery of configuration drift, and search performance.

        \subsection{Improvement Over the Baseline}{

            The iterative-depening depth-first search baseline approach searched for a working environment configuration using only the mutation operators \texttt{DecrementSemverMajor} and \texttt{DecrementSemverMinor}. We found that the most common baseline result was a timeout at the one hour mark without finding a working environment configuration. Search failed to complete for 978 out of 2,096 gists, nearly 50\%. 3,921 out of 5,423 notebooks likewise failed to complete within an hour. Comparatively, \feedback{} only timed out on 248 gists, a reduction of 730 from the baseline analysis. This is mirrored by Jupyter notebooks, only 488 of which timed out versus the baseline of 3,903, a reduction of 3,415.

            \insight{
                A simple search strategy is insufficient for fully exploring the configuration space of most code snippets. \Feedback{} aids exploration by allowing \Tool{} to focus only on a relevant subset of the configuration space.
            }

        }

        \subsection{Discovered Configuration Drift}{



            \Tool{} discovered 175 instances of configuration drift present in 143 gists and 73 instances present in 63 notebooks. Search terminated without finding a working environment configuration for 1,882 gists and 5,017 notebooks, only 29 and 24 of which timed out, respectively (although 1,839 notebooks experienced a Jupyter error related to network issues). For 928 gists and 1,710 notebooks, search was terminated because the validation failure was heuristically determined to not be fixable by mutating installed dependency versions according the criteria in Section~\ref{sec:v2:feedback:checkpointing}. In the other cases, search was terminated because \Tool{} had explored all environment configurations in the portion of the configuration space determined to be relevant to the validation failure without finding a patch.

            \insight{
                \Tool{} is capable of finding instances of configuration drift in both gists and notebooks. It can potentially find more than one instance in a single snippet.
            }

        }

        \subsection{Search Performance}{

            \begin{figure}
    \centering
    \includegraphics[width=\linewidth]{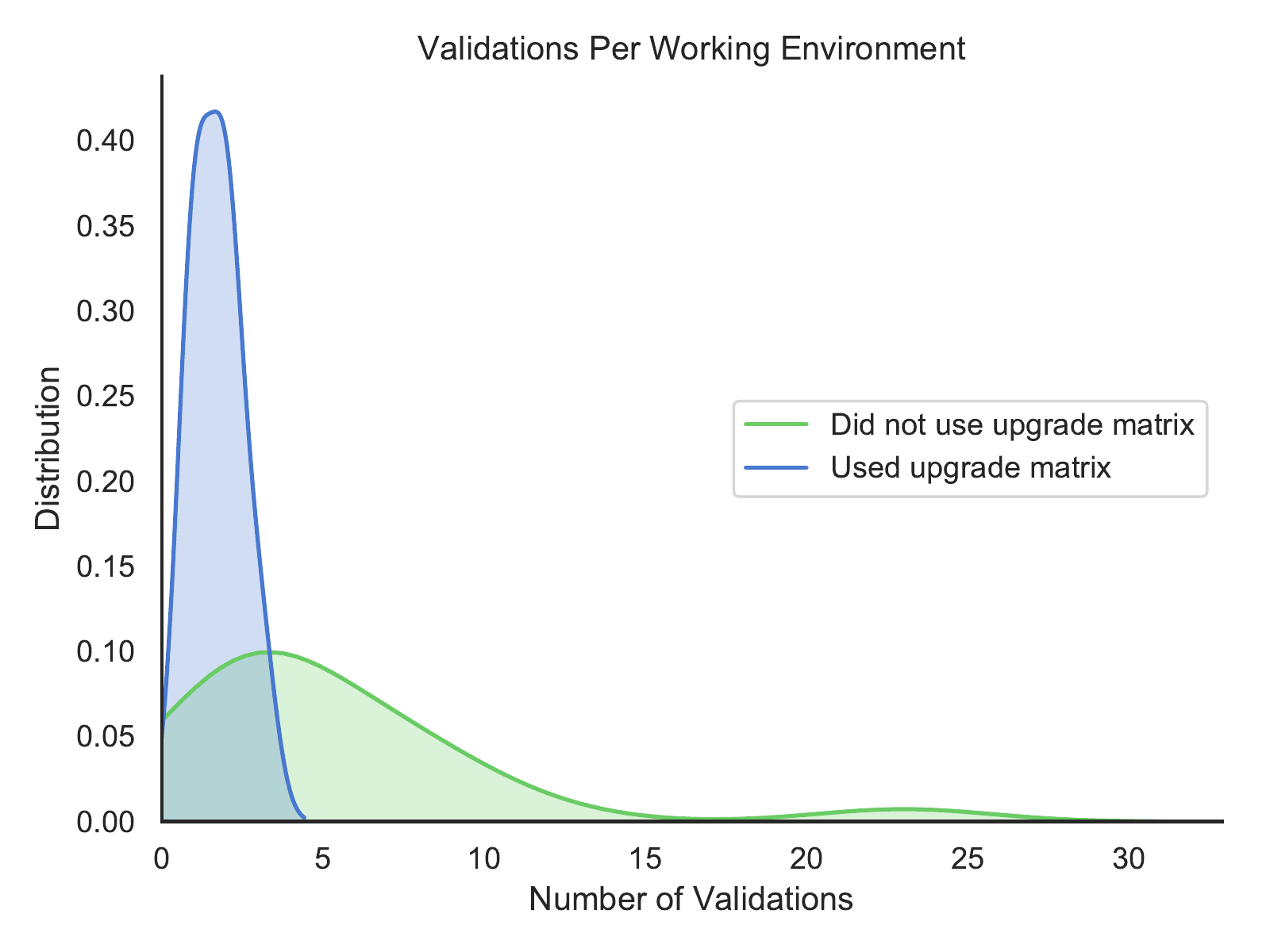}
    \caption{
        Number of validations performed on gists for which a working environment was ultimately found. If \Tool{} is able to make use of data in a version upgrade matrix, it can converge to a working environment with fewer validations.
    }
    \label{fig:successful-fixes}
\end{figure}

            We record two metrics important to search performance, corresponding to the objectives of \Search{} (Section~\ref{sec:v2:feedback}). The first is the total number of environment configurations validated before finding a patch. This indicates the overall performance of \feedback{}, since validation dominates runtime performance. The second metric is the number of mutations made to a candidate environment to find a patch. Both metrics reflect on \Tool{}'s ability to prune and prioritize within the search space.

            \Feedback{} shows an improvement over the baseline in the average number of validations needed to find a working environment configuration. Gists see a large improvement, dropping from 18 validations on average in the baseline to 8 when using \feedback{}. Jupyter notebooks see a more modest improvement from 7 to 6. For gists, \Tool{} was able to take advantage of a version upgrade matrix for 7 of the 27 (26\%) that were made fully executable, bringing the average number of validations needed down from 5 to 2. Even when a working environment was not found, \Tool{} was able to use the version upgrade matrices for about 28\% of code snippets, decreasing the number of validations needed before deciding no working environment existed.

            Figure~\ref{fig:successful-fixes} shows the distribution of number of validations required by \feedback{} to find a working environment configuration for gists. When able to prioritize using a version upgrade matrix, \Tool{} requires a third as many validations on average, with substantially better savings in the worst case. Time savings are also seen for individual instances of configuration drift, regardless of whether a fully working environment was eventually discovered.

            A relatively small number of mutations (about 2) is required to find a configuration patch for each instance of drift on average. However, incorporating information from a version upgrade matrix results in an improvement in the worst case. For gists, the maximum number of mutations made to create a configuration patch without a version upgrade matrix was 14, while the maximum with one was 7. For notebooks, the maximum number of mutations dropped from 11 to 4. This indicates that \Tool{} is able to correctly skip over certain versions while searching.

            \insight{
                \Feedback{} is able to find configuration drift by only searching a portion of the entire configuration space. Whenever a version upgrade matrix is available, \Tool{} can leverage the build information to prioritize dependency versions and further improve search performance.
            }

        }

        \subsection{Configuration Drift Found in Code Snippets}{

            We now detail \Tool{}'s behavior on two code snippets from our corpus, detailing how \feedback{} actually found and fixed instances of configuration drift.

            \subsubsection{Prioritization of Dependency Versions}{

                Consider the gist \mobi{},\footnote{\url{https://gist.github.com/5866756}} which contains the following imports from the Python library \texttt{Sphinx}.

                \begin{minted}{Python}
                    from sphinx import addnodes
                    from sphinx.builders.html import StandaloneHTMLBuilder
                    from sphinx.util.osutil import EEXIST, make_filename
                    from sphinx.util.smartypants import sphinx_smarty_pants as ssp
                \end{minted}

                Executing \mobi{} with the latest version of Sphinx (\texttt{2.0.1}), results in an import error for the module \texttt{sphinx\_smarty\_pants}. A developer might assume that the configuration is broken and look for a missing dependency that provides an extension to Sphinx. However, Sphinx extensions are provided from the \texttt{sphinx.ext} module, not the \texttt{sphinx.util} module. Another assumption is that a breaking change was made to the \texttt{smartypants} module.

                To validate this assumption, the developer must find a previous version of Sphinx for which \mobi{} executes successfully. They start by reverting from \texttt{Sphinx==2.0.1} to the previous major version \texttt{Sphinx==1.8.5}. However, executing with this version results in the same error. They change the major version again to \texttt{Sphinx==0.6.7}, a change which also requires using Python 2, but doing so results in an import error for \texttt{osutil} on the line before, indicating that the latest version change has actually broken a part of the configuration which previously worked.

                If a correct working version exists, it must be one of the other untested minor or patch level versions. Without any other information, discovering this version requires performing an exhaustive search though the \texttt{1.x} and \texttt{0.x} versions. Using a depth-first strategy, such as iterative-deepening depth-first search, the developer might additionally validate versions \texttt{1.7.9}, \texttt{0.5.2}, \texttt{1.6.7}, and \texttt{0.4.3} before finding a working configuration with version \texttt{1.5.6}.

                \Tool{}, by comparison, starts by validating Sphinx version \texttt{2.0.1} and encounters the original error, localizing it to the Sphinx module. Using the available upgrade matrix, it sees that the most recent version with upgrades that resulted in a broken build was \texttt{1.7.5}, and the only broken upgrade was from version \texttt{1.4.5}. It then mutates the environment, resulting in a configuration in which the gist successfully validates.

            }

            \subsubsection{Uncovering Multiple Instances of Configuration Drift}{

                \Tool{} is capable of uncovering more than one instance of configuration drift from the code snippets that it analyzes. For example, by using \feedback{}, \Tool{} demonstrates two instances of configuration drift in the gist \texttt{guess\_candidate\_model.py},\footnote{https://gist.github.com/eba2bf1a0ecd3e541146e98f35d49739} which the following excerpt is from.

                \begin{minted}{Python}
                    from sklearn.cross_validation import train_test_split
                    from keras.preprocessing import sequence, text
                    from keras.models import Sequential
                    from keras.layers import (Dense, Dropout, Activation, Embedding, LSTM, Convolution1D, MaxPooling1D)
                    ...
                    X = [x[1] for x in labeled_sample]
                    y = [x[0] for x in labeled_sample]
                \end{minted}

                This gist relies on two installable packages: \texttt{scikit-learn} and \texttt{Keras}. \Tool{} correctly generates candidate environment configuration for both Python 2 and Python 3 containing the latest versions of both of these packages.

                The first validation inside of a candidate environment results in a failure, returning the message ``\texttt{No module named `sklearn.cross\_validation'}'' even though the \texttt{sci-kit} package is included in the environment configuration. \Tool{} recognizes this as a potential instance of configuration drift and applies the mutation operator \texttt{DecrementSemverMinor(scikit-learn==0.20.3) -> 0.19.2}.

                The next validation also results in a failure, but this time with the message ``\texttt{No module named `tensorflow.'}'' \Tool{} determines that the previous failure has been fixed, because execution progressed past the line which caused the previous failure, and recognizes this as a potential instance of configuration drift involving \texttt{Keras}. In searching for a patch, it applies the mutation operators \texttt{DecrementSemverMajor(Keras==2.2.4) -> 1.2.2} and \texttt{DecrementSemverMajor(Keras==1.2.2) -> 0.3.3}.

                Finally, \Tool{} encounters the error ``\texttt{name `labeled\_sample' is not defined}.'' It again recognizes that the previous failure has been fixed. Further, it recognizes that this failure indicates an error with the gist that cannot be fixed by modifying the environment configuration and terminates search.

            }

        }

    }

    \section{Limitations}{

        \Tool{} presents a promising approach to finding and repairing instances of configuration drift within a code snippet and its dependencies. We show that it is capable of finding and repairing drift in a large corpus of open source Python code snippets. There are, however, some limiting factors which make it challenging to uncover all instances of drift.

        \paragraph{Silent Failures}{

            \Feedback{} requires that configuration drift result in a crash during snippet execution, because this failure is used to guide the search process and determine when the configuration drift has been patched. This requirement is sound for a large number of breaking API changes, such as removal of modules or a change in the code signature, because such changes result in runtime or compile time exceptions. However, some breaking changes in dependency behavior may cause a failure which does not manifest as a crash. \Tool{} cannot address such cases, although we note that the problem of determining if program behavior is expected is an open and challenging area of research~\cite{Barr:2015:Oracle}.

        }

        \paragraph{Mutation Operators and Ordering}{

            \Tool{} concentrates on major and minor dependency versions because they are the most likely to introduce or change behavior of an API, and its search strategy explores versions in the configuration space with equal precedence, only giving preference to configurations that are more similar to the original candidate configuration. However, developers differ in how they reason about the stability of dependencies, and communities differ on how they handle versioning~\cite{Bogart:2015:Stability}. Other search strategies may ultimately prove to be more effective at uncovering configuration drift.

        }

        \paragraph{Overzealous Pruning}{

            Parts of the \Tool{} search algorithm are informed by heuristics. In cases where a heuristic erroneously produces a false positive or false negative determination, search may be terminated prematurely. For example, if a validation failure is heuristically determined to not be fixable by modifying dependency versions, but a fix actually does exist, \Tool{} will stop exploration and be unable to report the configuration drift.

            For some dependencies, search is informed by a version upgrade matrix obtained from TravisCI build results. However, an upgrade matrix may be incomplete or not sufficiently represent some breaking upgrades, causing incorrect pruning.

        }

        \paragraph{Snippet Properties}{

            Some code snippets may be unsuitable for uncovering configuration drift. Gists can be overly general, relying on other services and requiring resources like authentication tokens~\cite{Horton:Gistable:2018}. Failures related to these resources may shadow configuration drift that could otherwise be discovered. Notebooks are heavily used within the data science community~\cite{Rule:2018}, and data science dependencies may be more stable than Python dependencies in general. Additionally, gists were restricted to having at most 10 direct dependencies. While this does not necessarily restrict the total number of dependencies, it may impact the distribution of results.

        }

        \paragraph{Discovery}{

            Because exploration of the configuration space is guided by feedback from validation, \Tool{} discovers at most one instance of configuration drift at a time. Further, validation can be an expensive process, and other instances of drift may be hidden by the most recent validation failure.

        }

    }

    \section{Discussion and Future Work}{

        Recent work has shown that managing dependencies is a large and time consuming problem developers face when engaging in configuration management~\cite{Seo:2014:BuildErrors, Sulir:2016, Urli2018, Horton:Gistable:2018, Horton:2019:DockerizeMe}. In particular, some developers have highlighted dependency management as one of the largest problems facing Python, requiring developers to spend non-trivial amounts of time on configuration for basic tasks.\footnote{\url{https://twitter.com/jeanqasaur/status/1104990612057518081}} This problem impacts data scientists, who must overcome the friction imposed by configuration just to perform their job.\footnote{\url{https://twitter.com/LibSkrat/status/1122857944675229698}}

        In addition, upgrading a version of a library that contains incompatibilities in data structures, signature changes to API calls, or behavior breaking changes~\cite{ruiz:2015} can result in additional effort to address these changes. If a developer fails to understand the nature of a dependency change or performs insufficient testing, they can introduce undetected faults in code. Such factors may cause a developer to become reluctant to update dependencies. But, if they delay updating code too long, they may be locked out of being able to use important new features only available in new versions, as it becomes more and more difficult to adapt their code. For this reason, detecting instances of configuration drift has direct implications for configuration management. Being able to highlight where code contains outdated API usages, even between dependencies of that code, can help developers locate and fix usage of APIs that have been changed, benefiting work in API migration. Further, by pairing each instance of drift with a corresponding fix, we motivate code repair~\cite{Weimer:2006:Patches}, and reuse. For example, repairing Jupyter notebooks can enable replicating important calculations even in older notebooks.

        Although we focus on Python, we believe detection of configuration drift is an important problem for other languages. For example, updating the MongoDB driver for Node.js from \texttt{2.0.x} to \texttt{2.1.x} affected the way order by parameters are used in the \emph{sort} function. Previously, parameters were allowed to be specified as list of objects: \verb|[{publish_date: -1}]| but in the newer version, list of lists must be provided: \verb|[["publish_date", -1]]|. \Tool{} is capable of detecting such a breaking change in any language so long as the code snippet may be parsed and validated, and previous versions of dependencies are made available.

        While the technique outlined in this paper shows promise, we believe there are several areas for improvement.

        \paragraph{Fuzzing}{

            Data in version upgrade matrices is necessarily limited, as it originates from upgrades to existing open source projects which both use upgrade bots and build through TravisCI. When data does exist for a package, the matrix is usually sparse. We believe that the amount and quality of available upgrade data can be augmented. One method would be to engage in fuzzing of dependency versions for libraries. For example, the upgrade matrix for \texttt{numpy} could be filled in by choosing a set of snippets or generated tests that cover the API and observing their execution with different versions of the library. This additional execution data could help determine clusters of incompatible versions that indicate a breaking change for particular API uses.

            Traditional fuzzing techniques can also be implemented to improve drift detection. Some code snippets, particularly those meant to be used as examples, define functions but do not execute them. Executing these functions with generated input would ensure greater coverage of the code snippet and its API uses, allowing further detection of instances where configuration has drifted.

        }

        \paragraph{Progression in the Face of Adversity}{

            \Tool{} currently only recognizes configuration drift caused by versions of installed dependencies. However, code snippets often depend on resources external to the configuration environment, such as communication with databases or REST APIs. When validation of a code snippet fails due to an external service, future work can focus on determining if the failure represents an instance of configuration drift and attempt to generate a fix. If a fix cannot be found, a ``mock'' patch sufficient to remove the validation failure could be inserted, allowing analysis of the snippet to continue. Missing resources, such as files on the local file system, could also be synthesized to allow code snippet analysis to continue.

        }

        \paragraph{Improvements to Search}{

            \Feedback{} uses heuristics and checkpointing to reduce the search space. This results in improved performance, but at the potential cost of completeness. This could be mitigated by still searching all versions, but using version matrices for prioritization. Additionally, the checkpointing process can be modified to allow for backtracking to a previous checkpoint if no solution is found (as might be the case in library version conflicts). However, \Tool{} does not currently do so.

            Finally, work can be done to improve the quality of search through the environment space. \Tool{} assumes that the original candidate environment is complete and contains all required dependencies, if not the correct versions. We leave it to future work how best to continue searching if it is discovered that this is not the case.

        }

        \paragraph{Optimization and Improvement}{

            The validation strategy employed by \Tool{} currently starts by creating and configuring a new environment. This is sufficient to guarantee correctness of validation, but requires an expensive configuration phase after every mutation performed by \feedback{}. We note that, because \feedback{} generates a new environment configuration by applying a single mutation to the previous configuration, \Tool{} could optimize the validation process by creating and continually mutating a single environment specification. This would greatly reduce the cost of validation.

        }

    }

    \section{Related Work}{

        Previous work has concentrated on evaluating the executability of code in the context of its configuration. Sul\'{\i}r and Porub\"{a}n find 38\% of Java builds fail due to dependency related errors~\cite{Sulir:2016}. McIntosh et al. agree that build maintenance imposes a large overhead on developers~\cite{McIntosh:2011}. Yang et al.~\cite{Yang:2016} evaluate Python code snippets, as do Horton and Parnin~\cite{Horton:Gistable:2018}. They find that Python code snippets are not executable by default due to configuration errors, and Horton and Parnin additionally highlight the difficulty of correct configuration, of which inferring correct dependency versions ranks highly.

        Additional studies have focused specifically on how developers manage dependencies and versions within environment configuration. Xavier, Hora, and Valente note that developers deliberately make breaking changes to APIs for several reasons related to code maintenance~\cite{Xavier:2017:APIs}. Further, developers struggle with different options for specifying dependency versions that are supported by many package management systems~\cite{Dietrich:2019:Versioning, Cito:DockerEcosystem}, and analysis of such systems motivates the need for better tooling to deal with problems related to versions~\cite{Decan:2017:OSS}.

        Although these works demonstrate that configuration of dependencies is a frequent cause of build and execution failure, and that developers have need of better tooling support of versions~\cite{Mirhosseini:2017:VersionBot, Xing:2007:Diff, Henkel:2005:CatchUp, Dagenais:2011:AdaptiveChanges}, no study attempts to uncover instances of configuration drift automatically. Horton and Parnin do classify 15 code snippets which have drifted and only execute successfully with an older version of a dependency (Table~\rom{2} from Gistable~\cite{Horton:Gistable:2018}), but this classification was performed by manually generating configuration scripts.

        Existing work has been performed on the automatic repair of environment configurations. Weiss et al. presented Tortoise, a system which synthesizes patches for configuration scripts, but can only do so by analyzing recorded developer actions~\cite{Weiss:Tortoise:2017}. Horton and Parnin work on inferring working environment configurations without the need of developer interaction, but they crucially ignore dependency versions~\cite{Horton:2019:DockerizeMe}.

        Work by Macho et al.~\cite{Macho:2018:BuildMedic} and by Hassan and Wang~\cite{Hassan:HireBuild} has presented automated repair of configuration scripts that includes version modifications. These are the closest related works of which we are aware. Crucially, both techniques require an existing environment configuration, and neither approach uses feedback to guide search. In addition, they focus on the Java build ecosystem, and cannot repair Python code snippets. Macho et al. prioritize versions using a distance metric that prefers small version changes. Hassan and Wang generate a ranked list of all possible patches and exhaustively apply them until some patch succeeds or there are no more patches to apply. Neither approach attempts to prune the configuration space, and prioritization is limited. There are approaches that use feedback~\cite{Nair:2018:FLASH} and historical knowledge~\cite{Le:2016:Repair}, but they focus on code, not computing environments.

        This work presents an improvement over previous approaches by incorporating feedback and prior knowledge for more efficient identification of repairs. We also focus on identifying instances of configuration drift in code, determining when dependencies have been updated, and synthesizing a patch as a proof of existence. This is in contrast to other approaches, which focus on an existing environment configuration and assume failures result from the code being updated.

    }

    \section{Conclusion}{

        Motivated by the problem of detecting out-of-date code snippets, we implement \Tool{}. \Tool{} discovers instances of configuration drift in Python code snippets, where each instance is identified by looking for crashes that can be fixed with modifications to versions of installed dependencies.

        Two techniques lie at the heart of \Tool{}'s ability to uncover configuration drift. The first is \feedback{}, a search strategy that uses information gleaned from executing a code snippet to determine which portion of the configuration space to explore next. The second is a corpus of version upgrade matrices that describe the before and after build statuses for over seven thousand package upgrade events in open source Python projects. These version upgrade matrices allow \Tool{} to significantly prune the configuration space, and effectively prioritize the remaining states, leading to faster discovery of working configurations.

        \Tool{} found 248 instances of configuration drift in the Gist and Jupyter datasets. It was able to do so quickly, requiring many fewer validations than a baseline prioritization strategy.
    }
    \section*{Acknowledgements}{\small
        This work is funded in part by the NSF SHF grant \#1814798.
    }

    \bibliographystyle{IEEEtran}
    \bibliography{main}

\end{document}